%% file: main.tex
\documentclass[aps,prd,physrev,superscriptaddress]{revtex4-2}

\usepackage{graphicx}% Include figure files
\usepackage{dcolumn}% Align table columns on decimal point
\usepackage{bm}% bold math
\usepackage{comment}
\usepackage{amsmath,amsfonts,amssymb}
\usepackage{url}

\begin{document}

\title{Instrumental development for Cryogenic sub-Hz cROss torsion bar detector with quantum NOn-demolition Speed meter (CHRONOS)}

\author{Daiki Tanabe}
\thanks{Corresponding author: tana2431.ts@gmail.com}
%\homepage[]{}
%\thanks{}
\affiliation{Institute of Physics, Academia Sinica, Taipei, Taiwan}
\affiliation{Center for High Energy and High Field Physics (CHiP), National Central University, Taoyuan, Taiwan}
\affiliation{Department of Physics, National Central University, Taoyuan, Taiwan}
\affiliation{Institute of Particle and Nuclear Studies, High Energy Accelerator Research Organization (KEK), Tsukuba, Japan}
%\affiliation{}

\author{Hsiang-Yu Huang}
\affiliation{Department of Physics, National Central University, Taoyuan, Taiwan}
\affiliation{Center for High Energy and High Field Physics (CHiP), National Central University, Taoyuan, Taiwan}

\author{Yuki Inoue}
\affiliation{Department of Physics, National Central University, Taoyuan, Taiwan}
\affiliation{Center for High Energy and High Field Physics (CHiP), National Central University, Taoyuan, Taiwan}
\affiliation{Institute of Physics, Academia Sinica, Taipei, Taiwan}
\affiliation{Institute of Particle and Nuclear Studies, High Energy Accelerator Research Organization (KEK), Tsukuba, Japan}

\author{Mario Juvenal S. Onglao III}
\affiliation{National Institute of Physics, University of the Philippines - Diliman, Philippines}
\affiliation{Department of Physics, National Central University, Taoyuan, Taiwan}
\affiliation{Center for High Energy and High Field Physics (CHiP), National Central University, Taoyuan, Taiwan}

\author{Ta-Chun Yu}
\affiliation{Department of Physics, National Central University, Taoyuan, Taiwan}
\affiliation{Center for High Energy and High Field Physics (CHiP), National Central University, Taoyuan, Taiwan}

\date{\today}

\begin{abstract}
Gravitational waves from intermediate-mass black-hole (IMBH) binaries is a probe of strong-field gravity and black-hole evolution. Detection of IMBH is challenging because of their typically low frequency where the seismic noise, radiation pressure noise, and thermal noise dominate. The Cryogenic sub-Hz cROss torsion bar detector with quantum NOn-demolition Speed meter (CHRONOS) has been proposed to reach a strain sensitivity of 10$^{-18}$~${\rm Hz}^{-1/2}$ at 2~Hz. It aims to detect GW from IMBH mergers with the mass of $\mathcal{O}(10^4)$~M$_{\odot}$ and to explore stochastic gravitational background of $\Omega_{\rm GW} \sim 2\times 10^{-3}$ at 2~Hz. We present the overview of the CHRONOS hardware which is designed to integrate key techniques for improving low frequency sensitivity; torsion bar, speed meter, and cryogenic mirror. As a demonstration of the interferometer operation, we also report the commissioning status of a Michelson interferometer in National Central University in Taiwan which has been assembled as a partial component of CHRONOS.
\end{abstract}

\maketitle

\input{1_intro}
\input{2_overview}
\input{3_commissioning}
\input{4_conclusion}

\begin{acknowledgments}
We would like to express our sincere gratitude to Y-C.Lin, M.Hasegawa, T.Kanayama and M.Hazumi for their valuable discussions and continuous support throughout this work.
We also acknowledge the support and collaborative environment provided by the Department of Physics and the Center for High Energy and High Field (CHiP) at National Central University, the Institute of Physics, Academia Sinica, the National Institute of Physics, University of the Philippines Diliman, as well as Taiwan Semiconductor Research Institute (TSRI).
Y.I. is supported by the National Science and Technology Council (NSTC) of Taiwan under Grant No. 114-2112-M-008-006, and by Academia Sinica under Grant No. AS-TP-112-M01.
\end{acknowledgments}

\bibliography{reference}

\end{document}

%% file: 1_intro.tex
\section{Introduction}\label{sec:intro}

Detecting gravitational waves (GWs) from binary black hole mergers heavier than 1000~M$_{\odot}$, those categorized in intermediate-mass black holes (IMBHs), is of high scientific interest for testing general relativity and unveiling black-hole evolution. The GW from heavier masses has its typical frequency at lower range where seismic noise, radiation pressure noise, and thermal noise dominates. To explore the low frequency GW, the Cryogenic sub-Hz cROss torsion bar detector with quantum NOn-demolition Speed meter (CHRONOS) has been proposed~\cite{CHRONOS}.

The CHRONOS is a Sagnac interferometer in which GW signals are reconstructed by measuring the velocity of the mirrors instead of their displacement. It deploys torsion-bar-shape test masses, Sagnac speed meter, and cryogenic mirrors to suppress the low-frequency noises. It aims to reach the strain sensitivity at the order of 10$^{-18}~{\rm Hz}^{-1/2}$ at 2~Hz. In addition to the observation of binary black holes, it aims to explore stochastic GW background of $\Omega_{\rm GW} \sim 2\times 10^{-3}$ at 2~Hz.

%% file: 2_overview.tex
\section{Hardware overview of CHRONOS interferometer} \label{sect:overview}

CHRONOS is a GW detector designed to observe the cross polarization of GW by measuring the rotational velocity of two orthogonally crossed torsion bars equipped with mirrors, using a Sagnac interferometer~\cite{CHRONOS}. In the final configuration, it has cross torsion bars with triangular cavities and dual recycling cavities~\cite{CHRONOS_optics}. Considering that the sensitivity depends on length of the triangle cavities, a configuration with 2.5-m length has been proposed as a prototype of CHRONOS. The size of the entire interferometer of the prototype is approximately 10-m square.

\begin{figure}[tb]
  \centering
  \includegraphics[width = 1.0\linewidth]{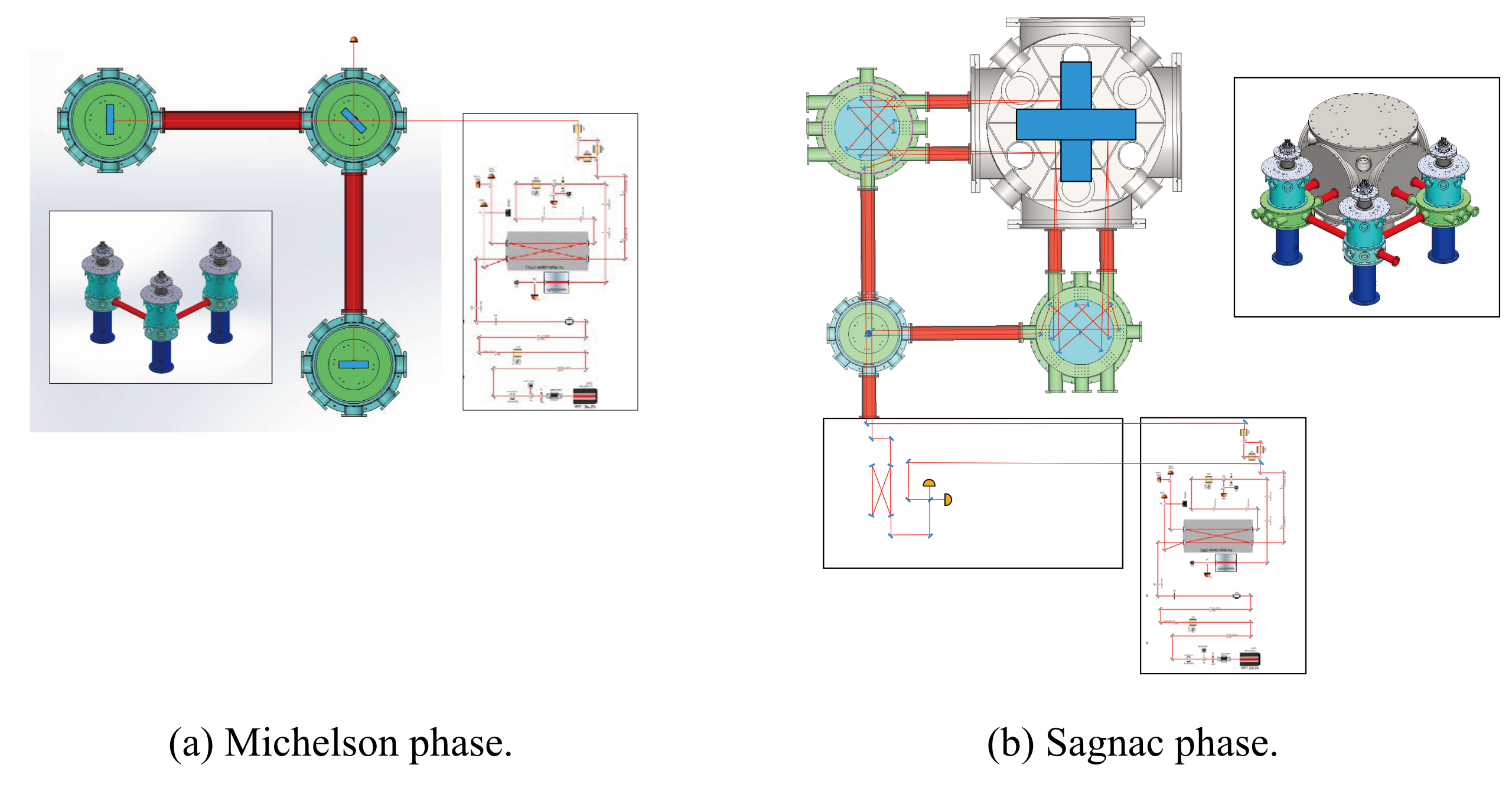}
  \caption{Interferometer configurations of CHRONOS commissioned in NCU during the Michelson and Sagnac phases.}\label{fig:phases}
\end{figure}

Toward the prototype configuration, it is gradually developed through the Michelson phase and the Sagnac phase as shown in Fig.~\ref{fig:phases}. We establish the basic techniques of suspension and input optics in the Michelson phase, and demonstrate speed-meter configuration with torsion bars in the Sagnac phase. The interferometer in these phases are commissioned in National Central University (NCU).

The final phase of CHRONOS has three key features; torsion bar, speed meter, and cryogenic mirror. Two bars equipped with mirrors are suspended from a pre-isolated tower for suppressing the seismic noise and to realize free-mass motion. Since the transfer function of the vibration from the suspension point to the mirror decreases with an inverse power law of frequency above the resonance, lowering the resonant frequencies enhances seismic isolation. The rotational mode of a torsion bar has a much lower resonant frequency than the pendulum modes. Consequently, with fibers of practically achievable length, the suspension can be designed to place the rotational resonant frequency in the millihertz region. Torsion bar has an additional advantage of canceling the laser intensity noise owing to the balance of torque at two mirrors on each bar~\cite{tanabe_chronos_intensity}. Focusing the low-frequency region also motivates us to adopt a photon-pressure actuator, which is insensitive to seismic and magnetic disturbances and does not require a recoil mass suspended nearby the test mass~\cite{tanabe_chronos_photon_pressure}.

A 1064-nm Nd:YAG laser is injected to each bar and split into two paths; clockwise and counterclockwise beams circulating in the triangular cavities. These paths acquire a differential phase shift induced by the bar’s rotation velocity. The resulting interference signal is read out at the output port with a balanced-homodyne method~\cite{speedmeter_homodyne}. Because velocity is in proportional to the inverse of frequency while displacement is in inverse square, speed meter can reduce the radiation pressure noise which rises in low frequencies. Balanced-homodyne detection enables us to subtract the common mode and control the detuning angles of power- and signal-recycling cavities independently from the homodyne readout angle.

To reduce thermal noise, the torsion bars used as test masses are cooled to 10~K using pulse-tube cryocoolers. Sapphire is chosen as the test-mass material because of its high thermal conductivity at cryogenic temperature~\cite{kagra_sapphire}. Because of the size limitations of industrially producible sapphire crystals, we fabricate each bar by bonding sapphire blocks using hydroxide-catalysis bonding~\cite{hydrocatalysis}. The vacuum chambers for the test masses are designed with three thermal layers at 300~K, 50~K, and 4~K.

Despite of the novel techniques combined in it, CHRONOS also deploys the common interferometry techniques which have been demonstrated in the conventional Michelson-type GW detectors represented by LIGO, Virgo, and KAGRA~\cite{aligo,virgo,kagra}. All mirrors in the interferometer are suspended to suppress the seismic noise and controlled by actuators to lock the interferometer. Parameter selections for control are described by Inoue {\em et~al.}~(2025)~\cite{CHRONOS_optics}. Laser is pre-stabilized at the input optics stage before injected into the interferometer.

%% file: 3_commissioning.tex
\section{Status of commissioning test} \label{sect:commissioning}

\begin{figure}[tb]
  \centering
  \includegraphics[width = 1.0\linewidth]{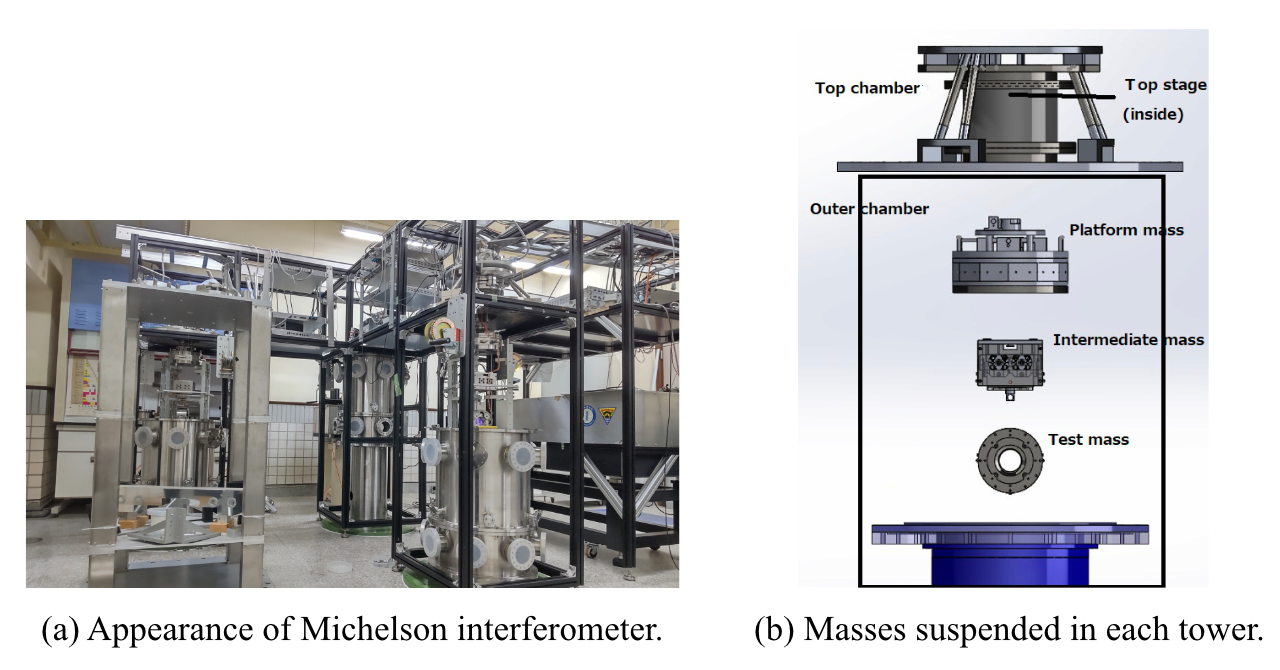}
  \caption{Michelson interferometer and its mirror suspension tower in NCU.}\label{fig:ncu_system}
\end{figure}

Since 2021, we have assembled one of the beam-splitter part of the CHRONOS interferometer in NCU. It forms a Michelson interferometer with arms of approximately 2-m length. Figure~\ref{fig:ncu_system}(a) shows the appearance of the interferometer in NCU. Using this Michelson interferometer, we have been conducting tests of feedback control of the mirror suspensions, pre-stabilization of the laser, and reconstruction of the arm-length strain from the interferometric signal. 

\begin{figure}[tb]
  \centering
  \includegraphics[width = 0.7\linewidth]{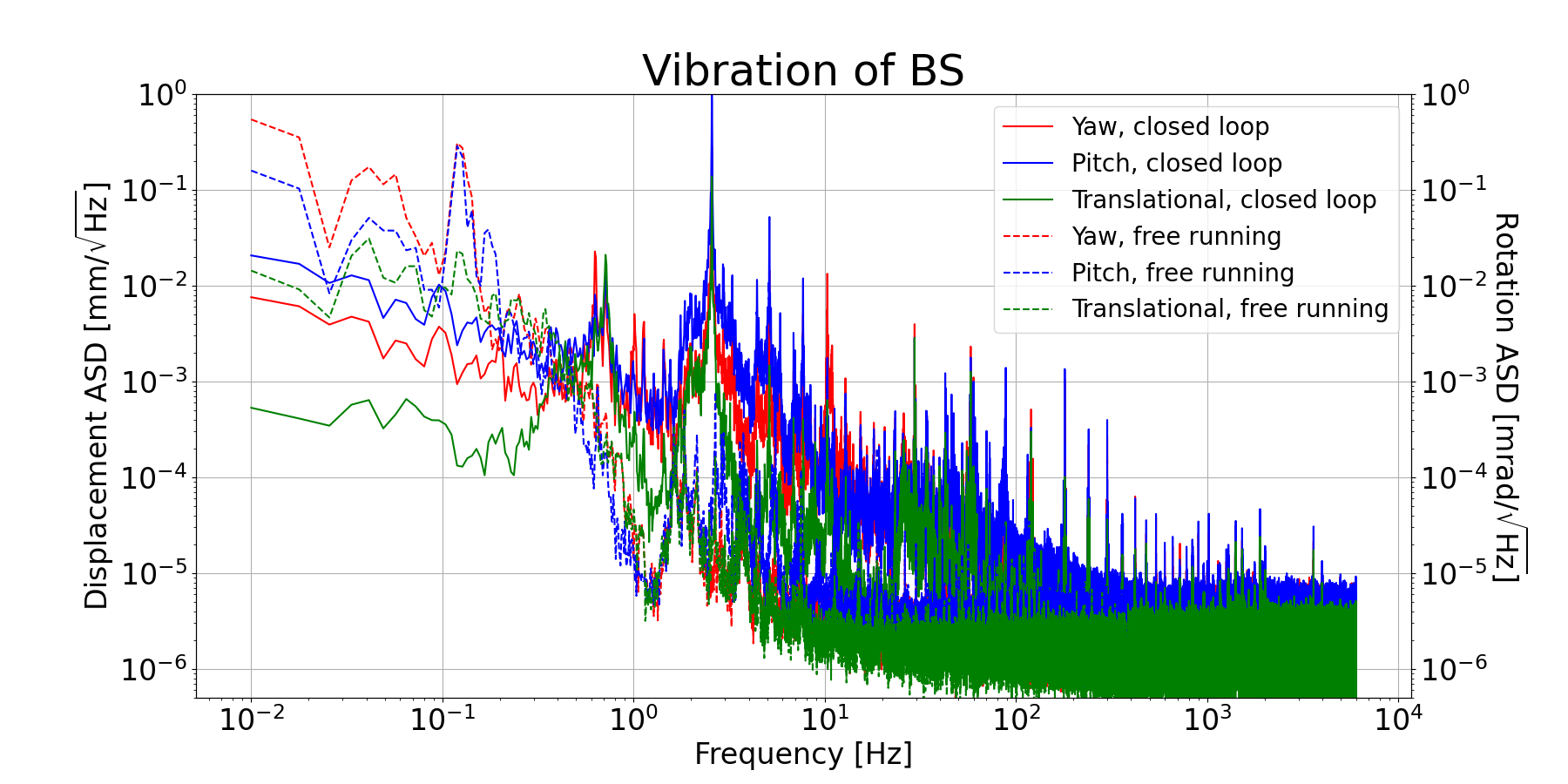}
  \caption{Noise spectrum of the motion of beam splitter. Unit is mm for translational displacement, while it is mrad for yaw and pitch rotation angle. {\it Red dash}: Yaw without feedback, {\it Blue dash}: Pitch without feedback, {\it Green dash}: Displacement without feedback, {\it Red solid}: Yaw with feedback, {\it Blue solid}: Pitch with feedback, {\it Green solid}: Displacement with feedback. }\label{fig:TM_spectrum}
\end{figure}

We are commissioning the control of the beam splitter (BS) in the center chamber. The BS and the mirrors are suspended through platform mass and intermediate mass as depicted in Fig.~\ref{fig:ncu_system}(b). We monitor the motion of BS by an optical lever with a 635-nm visible laser. The BS is controlled by coil-magnet actuator. Figure~\ref{fig:TM_spectrum} shows the noise spectra of the displacement, yaw, and pitch obtained with the closed-loop feedback control. It indicates stabilization of approximately -40 dB is achieved below 0.2~Hz.

\begin{figure}[tb]
  \centering
  \includegraphics[width = 0.7\linewidth]{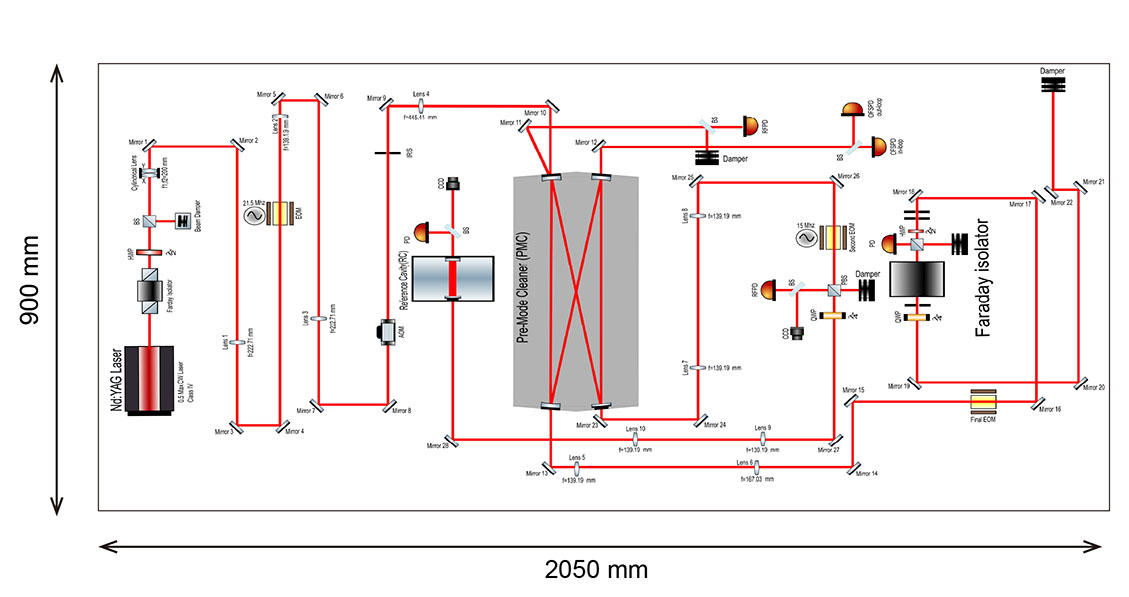}
  \caption{Schematic diagram of the CHRONOS input optics.}\label{fig:IO}
\end{figure}

Prior to injection into the interferometer, the main laser beam must be shaped to a pure Gaussian mode and stabilized. As shown in Fig.~\ref{fig:IO}, the input optics consists of stages for mode cleaning, intensity stabilization, and frequency stabilization. Selection of Gaussian mode is realized by locking the pre-mode cleaner (PMC), an aluminum bow-tie cavity. By controlling one of the PMC mirrors by a piezoelectric actuator, we maintained PMC lock for approximately one hour. A beam from other ports are used for intensity and frequency stabilization. By sending back the amplitude signal measured by a photodetector to an acousto-optic modulator, we improved of the relative intensity noise (RIN) by 20~dB at 10~Hz.

%% file: 4_conclusion.tex
\section{Discussion} \label{sect:discussion}
The motion of the test mass, locking time of PMC, and RIN achieved in this study have to be further improved. Displacement of each test mass has to be suppressed to be below 1~$\mu$m, while rotation should be below 1~$\mu$rad to achieve interferometer lock. We have to suppress the resonant peaks of yaw and pitch degree of freedom by feedback control. For further stability of the PMC lock, we have to deploy temperature control of the PMC because the optical path length non-negligibly changes by temperature. Temperature control of OFS could also improve RIN. Construction of the end test-mass chamber and Sagnac interferometer will follow the Michelson interferometer test.

\section{Conclusion} \label{sect:conclusion}
CHRONOS aims to detect sub-hertz GW by improving low-frequency sensitivity with torsion bar, speed meter, and cryogenic mirror. As a preparation step before Sagnac phase, We are conducting a commissioning test of Michelson phase to demonstrate fundamental operation techniques for interferometer. We achieved 40~dB vibration isolation of the test mass by feedback control at the frequency below 0.2~Hz. In the input optics, we achieved to lock the PMC for 1~hour and suppress RIN for 20~dB at 10~Hz.